# SENSORY SUBSTITUTION : PERCEPTION DEDICATED TO ACTION
Cognitive processes involved in the utilization of a simple visuo-tactile sensory prosthesis

Hardy B.[1], Ramanantsoa M.-M.[1], Hanneton S.[12], Lenay C.[2], Gapenne O.[2] and Marque C.[2]

[1] Université Paris 5 René Descartes, UFR STAPS, Laboratoire des Sciences du Sport, Equipe Cognition et Motricité 1, rue Lacretelle 75015 Paris (France)
[2] Université de Technologie de Compiègne, Département TSH, Equipe Costech. BP 60.319, 60206 Compiègne CEDEX (France)

Paul Bach Y Rita [1] is the precursor of sensory substitutions. He started thirty years ago using visuo-tactile prostheses with the intent of satisfying blind people. These prostheses, called Tactile Vision Substitution Systems (TVSS), transform a sensory input from a given modality (vision) into another modality (touch). These new systems seemed to induce quasi-visual perceptions. One of the author's interests dealt with the understanding of the coupling between actions and sensations in perception mechanisms [4]. Throughout his search, he noticed that the subjects had to move the camera themselves in order to recognise a 3D target-object or a figure placed in front of them. Our work consists in understanding how sensory information provided by a visuo-tactile prosthesis can be used for motor behaviour. In this aim, we used the most simple substitution device (one photoreceptor coupled with one tactile stimulator) in order to control and enrich our knowledge of the ties between perception and action.

This study rested on two purposes:
- one -more technical- consisted in proving the device was working;
- the other- more theoretical- concerned the coexistence of two modes in the prosthesis utilisation.

Indeed, in the visuo-manual co-ordination paradigm, it is useful to segregate two types of representations of perceived stimuli : semantic and pragmatic ones [2][3] showed that pragmatic representation established a direct connection between sensory stimulus and motor response. However, action could also rely on a semantic system once verbal response required to built a semantic representation of the target location. Our topic focused on this separation and we attempted to observe whether or not it could be demonstrated in an experimental context involving a perception substitution system. Moreover, we were looking for possible differences between normal and blind people.

## METHODS

### Subjects
Thirty participants were separated into three groups: twenty normal subjects who were blind, ten were pointing in "direction" condition and ten in "distance" condition, and in the third group ten blind people were tested only in " direction condition" (see below for the description of distance and direction conditions).

### Procedure
Each subject wore a prosthesis on the left index finger consisting in a photo-transistor driving a tactile stimulator. The tactile stimulator was active when light measured by the cell was beyond a fixed threshold.

### Apparatus
All participants were seated in front of a sagittal panel hidden from their view (for the first and second group), with the left hand (wearing the prosthesis) resting on the table and the right hand on a computer mouse that sent back tactile vibrations (starting position). When the prosthesis was moved towards a luminous source, the subject was informed by a continuous vibration.

Before starting the test, subjects participated in a training program consisting in a passive learning of the six target locations. The learning phase ended when subjects were able to reach all targets in random order (with an distance error inferior to 1 cm).

Target locations and experimental conditions
- **Condition 1** ("direction" condition): one array of six targets in the shape of an arch centred on the starting position ;
- **Condition 2** ("distance" condition): one array in the shape of a line crossing the starting position.

Each subject ran 60 pointing trials. When he "localized" the target with the prosthesis (left hand), he was instructed by the examiner to make a "go" pointing or a "chiffre" pointing with the right hand. In the first case he pointed while saying "go", and in the second case, he pointed while he was pronouncing the number corresponding to the estimated target location. The 60 trials were interrupted every 20 trials by a repeat of the passive memorisation of the location of targets.

Data analysis
Three measures were considered:
- the **variable error** which gives an estimation of the pointing precision in accounting for average distance between finger's pointing and target's position. This first variable was used to measure the ability of subjects to distinguish the different target locations.
- The **constant error** was used to check typical seek landmark. Precisely we were looking for a systematic deformation between the "actual space" and subjects' perceptual space.
- Lastly, we analysed also the **percentage of correct answers** (verbal answers). Pointing results can also be converted into p.c.a by considering a "pointed answer" corresponding to the number of the target closest to the pointed location.

Hypotheses
Three hypotheses can be considered :
1. a pragmatic system manages "go" pointing condition while a semantic system deals with "chiffre" pointing and verbal categorisation [6].
2. the whole experience was driven by a semantic system.
3. a semantic system emerges only from verbal categorisation and a pragmatic one drives the "chiffre" and "go" pointing.

## RESULTS

First, we wanted to prove our new prosthesis was working. From the confidence ellipses, we observed that subjects were able to distinguish between the six targets. The mean variable error was 5.4 cm for sighted subjects and 4.8 for blinds (the difference was not significant, $F[1, 108]= 3.541$ $p<0.063$). Confidence ellipses were smaller for blind subjects than for normal subjects. However, accuracy in the "direction condition" for targets 1 and 2 was significantly better for blind people than for sighted one.

In the "direction" condition, the major axis of confidence ellipses is quasi-perpendicular to the direction of targets. In the "distance" condition, this axis is quasi-parallel to the alignment of targets (figure 1). According to the results obtained by Rossetti and al. in normal pointing tasks [5][6], the relative orientations of confidence ellipses give strong indications to favour the second hypothesis : the pointing answers of subjects seems to be driven by a semantic system.

Finally, the analysis of the percentage of correct answers shows that there was a strong correlation between the three possible situations ("pointed answer" in "go" condition, pointed answer in "chiffre condition" and verbal answer in "chiffre" condition). In the "chiffre" pointing condition, we showed that there was a strong dependency between the probability to give a correct verbal answer and the probability to give a correct "pointed" answer ($F[1,18]=48,32$, $p<0.000002$). The verbal and pointed responses are driven by the same (semantic) discrimination process.

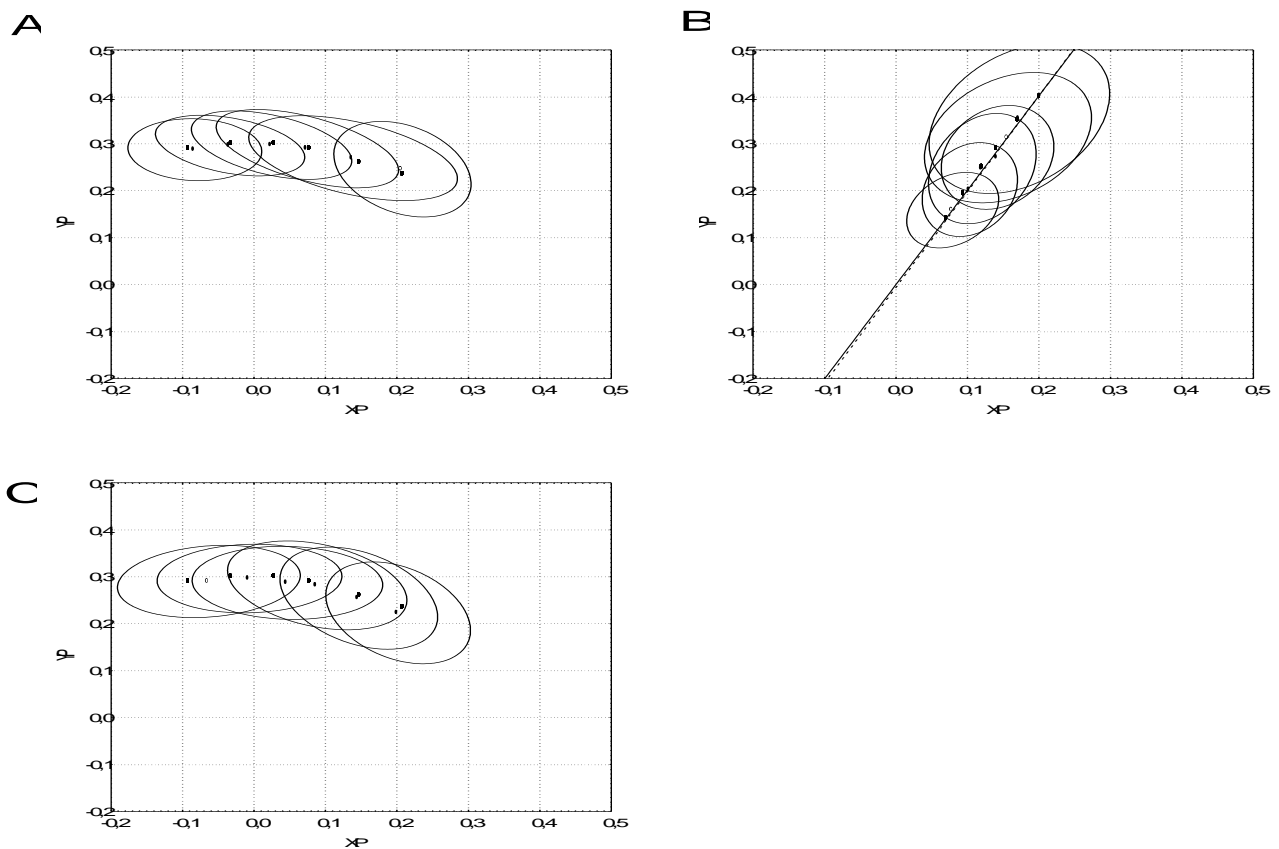

**Figure 1. Confidence ellipses for the direction and distance conditions (A and B, C = blind people)**

## DISCUSSION

Results are inclined to favour the second hypothesis. There was no significant differences in the pointing results between the "chiffre" and "go" conditions, in direction as in distance and by the whole groups. Moreover, the orientation of ellipses showed that the location of each target was evaluated relatively to the other targets (semantic mode). This result appeared without any differentiation in the two conditions "chiffre" and "go" by blind people. That would certify a semantic processing in the two pointing conditions. Indeed, we propose that our prosthesis was used only in a semantic way by subjects. This semantic mode may be forced by the experimental paradigm, and particularly, by the absence of a learning phase allowing a calibration of sensory-motor perceptual space. This calibration can occur only if subjects are allowed to determine the target location themselves and to explore it with the prosthesis. This direct manipulation of targets by subjects allows the closure of the "action/perception loop" from which the constitution of a perception space originates. This is not the case here: in our experimental conditions, subjects did not learn a perceptual space but only the location of the targets in a passive way (mostly a proprioceptive way ?). The prosthesis was used by subjects only to discriminate the targets and then to choose in his memorised repertoire the correct verbal or pointed answer. Action (pointing) was not directly driven by perception, but eventually by a memorised proprioceptive representation of the answer.

The next step for this research is obviously to try to provoke the emergence of a pragmatic use of the prosthesis by adding before the test a learning period. During this pre-test learning period, subjects will be instructed to manipulate the target themselves and to explore it with the prosthesis. This period will allow the closure of the action/sensation loop which is, on our opinion, necessary to the constitution of this kind of assisted perception. The study of cognitive and sensory-motor processes involved in the perception assisted by prostheses represents a perspective of reflection and fruitful research, especially to guide the learning and acceptance of prostheses by visually impaired people [1][3].